\documentclass[twocolumn,showpacs,preprintnumbers,amsmath,amssymb,superscriptaddress,aps,standalone,floatfix]{revtex4-1}
\usepackage{graphicx}% Include figure files
\usepackage{dcolumn}% Align table columns on decimal point
\usepackage{bm}% bold math
\usepackage{braket}
\usepackage{textcomp}
\usepackage{gensymb}
\usepackage{verbatim}
\usepackage{amsmath}
\usepackage{color}
\begin{document}

%\preprint{APS/123-QED}

\title{Determination of local defect density in diamond by double electron-electron resonance}

\author{Shang Li}
\thanks{These authors contributed equally to this work.}
\affiliation{Department of Physics, Graduate School of Science, Nagoya University, Nagoya 464-8602, Japan}

\author{Huijie Zheng $^*$}
%\thanks{These authors contributed equally to this work.}
\email{zheng@uni-mainz.de}
\affiliation{Johannes Gutenberg-Universit¨at Mainz, 55128 Mainz, Germany}

\author{Zaili Peng}
\thanks{These authors contributed equally to this work.}
\affiliation{Department of Chemistry, University of Southern California, Los Angeles, California 90089, USA}

\author{Mizuki Kamiya}
\affiliation{Department of Physics, Graduate School of Science, Nagoya University, Nagoya 464-8602, Japan}

\author{Tomoyuki Niki}
\affiliation{Department of Physics, Graduate School of Science, Nagoya University, Nagoya 464-8602, Japan}

\author{Viktor Stepanov}
\affiliation{Department of Chemistry, University of Southern California, Los Angeles, California 90089, USA}

\author{Andrey Jarmola}
\affiliation{Department of Physics, University of California, Berkeley, California 94720, USA}
\affiliation{ U.S. Army Research Laboratory, Adelphi, Maryland 20783, USA }   

\author{Yasuhiro Shimizu}
\email{yasuhiro@iar.nagoya-u.ac.jp}
\affiliation{Department of Physics, Graduate School of Science, Nagoya University, Nagoya 464-8602, Japan}

\author{Susumu Takahashi}
\email{susumu.takahashi@usc.edu}
\affiliation{Department of Chemistry, University of Southern California, Los Angeles, California 90089, USA}
\affiliation{Department of Physics and Astronomy, University of Southern California, Los Angeles, California 90089, USA}

\author{Arne Wickenbrock}
\affiliation{Johannes Gutenberg-Universit¨at Mainz, 55128 Mainz, Germany}
 \affiliation{Helmholtz-Institut, GSI Helmholtzzentrum f{\"u}r Schwerionenforschung, 55128 Mainz, Germany}

\author{Dmitry Budker}

\affiliation{Johannes Gutenberg-Universit¨at Mainz, 55128 Mainz, Germany}
\affiliation{Department of Physics, University of California, Berkeley, California 94720, USA}
 \affiliation{Helmholtz-Institut, GSI Helmholtzzentrum f{\"u}r Schwerionenforschung, 55128 Mainz, Germany}

%\thanks{}

\date{\today}

\begin{abstract}
Magnetic impurities in diamond influence the relaxation properties and thus limit the sensitivity of magnetic, electric, strain, and temperature sensors based on nitrogen-vacancy color centers. Diamond samples may exhibit significant spatial variations in the impurity concentrations hindering the quantitative analysis of relaxation pathways. Here, we present a local measurement technique which can be used to determine the concentration of various species of defects by utilizing double electron-electron resonance. This method will help to improve the understanding of the physics underlying spin relaxation and guide the development of diamond samples, as well as offering protocols for optimized sensing.
\end{abstract}

%\pacs{74.70.Xa, 71.30.+h, 74.25.Dw, 76.60.Cq, 76.60.Es}	
%\keywords{}

\maketitle


\section{Introduction}

Negatively charged nitrogen-vacancy (NV$^{-}$) color centers in diamond are at the core of various modern sensors that combine high sensitivity with spatial resolution down to the atomic scale~\cite{Gruber1997, schirhagl2014nitrogen}.
Because of NV$^{-}$'s long ground-state spin-coherence time and the ability to prepare and detect the spin state optically at ambient conditions~\cite{Jelezko2004, Childress2006, Epstein05, Gaebel06, togan2010, robledo2011high}, NV$^{-}$-based sensors find use in a wide range of applications such as readout and storage of information, nanoscale magnetic detection and imaging~\cite{Degen08, Maze2008, Balasubramanian2008, mcguinness2011, kehayias2017solution, barry2016optical, devience2015nanoscale, shi2015single}, temperature~\cite{Acosta2010}, pressure~\cite{doherty2014electronic}, electric-field~\cite{dolde2011electric, doherty2014electronic} sensing, as well as quantum computing~\cite{popkin2016quest}.

In order to improve the performance of NV$^{-}$ centers and make their applications practical, it is important to understand their fundamental properties such as the longitudinal and transverse relaxation times $T_{1}$ and $T_{2}$ \cite{mrozek2015longitudinal, jarmola2012temperature, jarmola2015longitudinal, stepanov2016determination, Takahashi08}.
A key to understanding the relaxation processes is to know the local environment around NV$^{-}$ spins, which includes magnetic impurities such as single substitutional nitrogen impurities (the so-called P1 centers), $^{13}$C, charge-neutral NV centers (NV$^{0}$), as well as neighboring NV$^{-}$ centers.
In previous works~\cite{wang2013spin, sasaki2016magnetic, bauch2020decoherence,wood2016wide}, measurements of the spin-coherence evolution were used to determine local concentrations of P1 spins.
For example, a recent study showed that 1/$T_{2}$ exhibits a linear dependence on the P1 concentration ([P1]) at [P1] $\geq$ 0.5 ppm, indicating that interactions with P1 bath spins are the dominant source of the spin decoherence of NV$^{-}$ centers~\cite{bauch2020decoherence}.
Another study~\cite{choi2017depolarization} suggests that longitudinal evolution rates can be used to analyze the defect density.
However, these methods are insufficient to distinguish the contribution of different defect species when more than one of them may play a role. 
In contrast, a method based on double electron-electron resonance (DEER) enables determining the concentrations of a selected group of spins. 
A DEER-based spin concentration measurement has been demonstrated in conjunction with high-field electron spin resonance (HF-ESR) to determine a wide range of the bulk [P1] in diamond and to study the dependence of $T_{2}$ on [P1]~\cite{stepanov2016determination}.
More recently, a DEER-based spin concentration measurement was used in combination with optically detected magnetic resonance (ODMR) spectroscopy to characterize the concentrations of NV$^{-}$ and P1 spins as part of the optimization of NV-ensemble formation in diamond layers~\cite{Kucsko18, Eichhorn19}.

In this study, we use optical DEER measurements with NV$^{-}$ centers to determine the local concentrations of paramagnetic spins in single crystal diamond samples.
In particular, the present study focuses on the characterization of the diamond crystals with a high spatial inhomogeneity of the local concentration.
As an extension, this method can be used to determine the local concentrations of various impurity spins in diamond as well as spin systems located outside the diamond without an additional reference sample.
We describe the method based on an experimental approach and theory presented in Ref.~\cite{stepanov2016determination}, then discuss the results of our measurements on several samples. These samples were used in earlier studies relevant to sensing based on dense ensembles of NV$^{-}$ centers. 

\section{Experiment}
\begin{table}[b]
\caption{\label{tab:DEERResult}Properties and treatment of the HPHT diamond samples used in this work.}
\begin{center}
\begin{tabular} {|c |c |c | c | }
\hline
Sample & Orientation & E-beam energy, dose & Annealing \\
\hline \hline
S2 	& [111] 	& 3 MeV, $10^{18}$ cm$^{-2}$ 	& 1050 $^\circ$C, 2 hrs \\
S5 	& [100] 	& 3 MeV, $10^{18}$ cm$^{-2}$ 	& 1050 $^\circ$C, 2 hrs \\
D12 	& [100] 	& 14 MeV, $10^{18}$ cm$^{-2}$ 	& 700 $^\circ$C, 3 hrs \\
F32 	& [100]	& 14 MeV, $10^{18}$ cm$^{-2}$ 	& 700 $^\circ$C, 3 hrs \\ 
E6 	& [100]	& 4.5 MeV, 5.9$\times10^{18}$ cm$^{-2}$ 	& 800 $^\circ$C, 16 hrs \\ \hline
\end{tabular}
\end{center}
\end{table}
The experiments were performed on five single-crystal diamond plates named S2, S5, D12, F32 and E6 (see Tab.\,\ref{tab:DEERResult}).
All samples are high-pressure, high-temperature (HPHT) synthesized diamond crystals.
The D12, F32, E6, and S5 samples are [100]-cut, and the S2 sample is [111]-cut diamond crystals. 
In order to increase the NV$^{-}$ concentration in the sample crystals, electron-beam irradiation was performed. 
For example, S2 and S5 samples were irradiated with 3 MeV electron beams at a dose of 10$^{18}$ cm$^{-2}$.
After the electron beam irradiation, the samples were thermally annealed at 1050 $^{\circ}$C in a forming gas (0.96 atm Ar and 0.04 atm H$_2$) for two hours.
The D12 and F32 samples were irradiated with 14 MeV electrons at a dose of 10$^{18}$ cm$^{-2}$ and annealed at 700 $^{\circ}$C for three hours. In addition to the inhomogeneity of the nitrogen concentration, the irradiation is spatially nonuniform for S5 and F32.
Details of the sample preparation process were reported previously~\cite{acosta2009diamonds}.
According to the previous characterization, the initial nitrogen impurity concentrations were $\lesssim 100\ $ppm and $\lesssim 200\ $ppm for S2 (Sumitomo) and S5 (Element Six), respectively.
After the NV$^{-}$ fabrication process the concentrations of NV$^{-}$ centers ([NV$^{-}$]) were estimated to be 16 ppm and 12 ppm for the S2 and S5 samples \cite{acosta2009diamonds}.

\begin{figure}[t]
\includegraphics[width=8 cm]{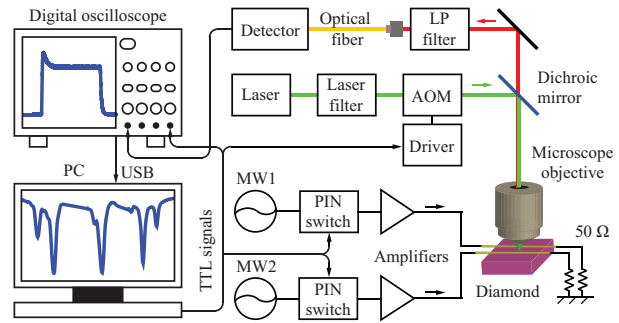}
\caption{\label{fig:diagram} Overview of the ODMR setup at Berkeley.
Two microwave signal generators (MW1 and MW2) are connected to  electronic microwave switches and amplifiers.
The switches are controlled with transistor-transistor logic (TTL) signals generated with a pulse-generator board mounted in a personal computer (PC).
The microwave wires are attached to the diamond and terminated with 50 ohm resistors.
The excitation light pulse is generated with a laser beam passing through an acousto-optic modulator (AOM).
A laser filter is used to reject an unwanted range of wavelengths from the laser.
The excitation laser beam is focused with a microscope objective, which is also used to collect the photoluminescent (PL) light from NV$^{-}$ centers. 
The PL signals are filtered with a dichroic mirror and long-pass (LP) filter and detected with an avalanche photo-detector.
An optical fiber is employed for confocal PL imaging.}
\end{figure}
In the present investigation, the optically detected magnetic resonance (ODMR) experiments were performed at Berkeley, Nagoya and Mainz.
An essentially similar schematic of the ODMR setup used in all three laboratories is shown in Fig.\,\ref{fig:diagram}.
The apparatus is a conventional confocal ODMR setup in which the NV$^{-}$ centers are illuminated with laser light (532 nm) pulsed with an AOM.
The laser light is focused on the diamond sample with a microscope objective.
The size of the laser spot is $\approx10\,\mu$m in diameter.
The red and near-infrared photoluminescence (PL) of NV$^{-}$ centers is collected through the same microscope objective and is separated from the excitation laser light with a dichroic mirror and a long-pass filter.
Then the PL is focused on a 62 $\mu$m diameter optical fiber for spatial filtering and is detected with an avalanche photo-detector .

\section{Theory}\label{sect:theory}
Here we discuss a model to analyze DEER signals and determine the concentration of target spins.
The DEER signal originates from a change of the NV$^{-}$ coherent state due to a shift of the magnetic dipole field from target spins with the application of the DEER pulse,
therefore the intensity of the DEER signal is related to the strength of the dipole field felt by the NV$^{-}$ centers (which is a function of the spin concentration) as well as the flip rate of the target spins (i.e., a change in the spin projection of the target spins) during the DEER measurement.

We first estimate the probability of spin flip events in the experiment.
The spin-spin dipolar interaction is a major source of spin-echo (SE) decay in solids.
Because the SE decay is affected differently by spins depending on whether they are excited by microwaves or not, spins under study are often divided into two groups; the excited spins (A spins) and the unexcited spins (B spins).
Here we consider spectral diffusion, which is a common process for SE decay for dilute spins in solids, including P1 and NV$^{-}$ spins in type-Ib diamond crystals \cite{Takahashi08, stepanov2016determination}.
In the spectral diffusion process, fluctuations of B spins (such as random spin flip-flops) diffuse the resonance frequency of A spins via the dipolar interactions.
According to a previously used model~\cite{Salikhov1981}, the SE decay due to spectral diffusion is described by
\begin{equation}
\label{eq:a}
\begin{split}
& SE(2\tau, W_{max}) = \\ 
& \exp\left(
- n \int\displaylimits_{0}^{\infty} f(W,W_{max}) \int\displaylimits_{V} \big[1 - v_0(2\tau, r, \theta, W)\big] \, \text{d}V \, \text{d}W
\right),    
\end{split}
\end{equation}
where $W$ is the flip rate of the bath spins, $n$ is the spin concentration of B spins, $f(W,W_{max})$ is the distribution function of the spin-flip rate with the peak at $W=W_{max}$~\cite{Salikhov1981},
\begin{equation}
\label{eq:fww}
f(W,W_{max}) = \sqrt{\frac{3 W_{max}}{2 \pi W^3}} \exp{\left(- \frac{3 W_{max}}{2 W}\right)},
\end{equation}
and $v_0(2\tau,r,\theta,W)$ represents the SE signal of a single A spin dipolar-coupled to a B spin with the relative radius vector $\vec{r}(r,\theta)$, given by
\begin{equation}
\label{eq:vtau}
\begin{split}
& v_0(2\tau, r, \theta, W) = \\ 
&\Bigg{[} \left(\text{cosh} (R(r, \theta, W)\tau ) + \frac {W}{R(r, \theta, W)} \text{sinh} (R(r, \theta, W) \tau) \right)^2 \\
& + \frac {A(r, \theta)^2}{4 R(r, \theta, W)^2} \text{sinh}^2 \left(R(r, \theta, W) \tau\right) \Bigg{]} \, \exp{\left(- 2 W \tau\right)}.
\end{split}
\end{equation}
$A(r, \theta) = \mu_0 \mu_B^2 g_A g_B(1 - 3 \cos^2 \theta)/(4 \pi \hbar r^3)$ represents the frequency shift due to the dipolar interaction between A and B spins and $R(r, \theta, W)^2 = W^2 - \frac{1}{4}A(r, \theta)^2$, where $\mu_0$ is the vacuum permeability, $\mu_B$ is the Bohr magneton, $\hbar$ is the reduced Planck constant, $g_A \sim g_B \sim 2$ are $g$-factors of the excited and bath spins, respectively.
The integration over the sample volume $V$ in Eq.~(\ref{eq:a}) takes into account all possible distances ($r$) and angles ($\theta$).
Therefore, the SE decay due to spin diffusion depends on the spin concentration ($n$) and the spin flip rate ($W$),
i.e., both stronger dipole interaction and higher spin flip-flop rate shorten the SE decay time.
Finally, using Eq.\,\eqref{eq:fww}, the average spin flip rate ($\langle W \rangle$) is given by $\sim 7.1$ $W_{max}$ as shown previously~\cite{stepanov2016determination}.
Thus, the probability of spin-flip events in a time evolution of $2\tau$ (2$\tau\langle W \rangle$) is given by $\sim 14.2 \tau W_{max}$.
The DEER model described below is considered based on the assumption that the spin-flip events are negligible.
The condition is validated by analyzing experimentally determined $W_{max}$ and the spin-concentration results (see Sect.~\ref{analysis}).

\begin{figure}[b]
\includegraphics[width = 70 mm]{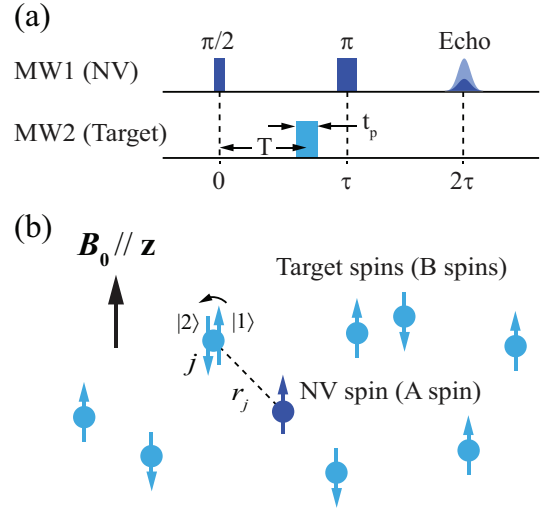}
\caption{\label{fig:DEERtheo} Overview of DEER model.
(a) A pulse sequence of DEER experiment. 
In DEER measurements of NV ODMR, a third MW1 pulse ($\pi/2$-pulse) is applied to map the echo intensity to the population of the NV's $m_S=0$ state.
The third pulse is not shown in this sequence.
(b) A schematic to illustrate the change of the magnetic dipole coupling between a NV and j-th target spins caused by a DEER pulse with a length of $t_p$.
}
\end{figure}
We next describe the DEER intensity when the spin flip process is negligible. 
The DEER intensity is used to determine the spin concentration ($n$) of B spins from the experimental DEER signals.
Figure~\ref{fig:DEERtheo} shows a pulse sequence of the DEER experiment consisting of the spin echo sequence of MW1 for NV centers and a DEER pulse of MW2 for excitation of target spins.
The DEER pulse to flip the target spins induces a shift of the magnetic dipole field at the NV causing phase shifts of the NV's coherent state.
For $t_p \ll 2\tau_{1}$, the accumulated phase shift during 2$\tau$ is expressed by,
\begin{equation}
\begin{split}
\label{eq:f}
&\delta \varphi = \frac{g_A \mu_B}{\hbar} \sum_{j} [ b_j (T - t_p/2) \\
&+ b_j^{MW}(t_p) \{(\tau - T - t_p/2) - \tau\} ],
\end{split}
\end{equation}
where $g_A$ is the g-factor of the NV center, $\mu_B$ is the Bohr magneton, $\hbar$ is the reduced Planck constant, $b_j \equiv \mu_0 \mu_B g_B (3 \cos^2 \theta_j - 1) \sigma_j /(4 \pi \hbar r_j^3)$ is a magnetic field produced by the $j$-th B spin at the NV spin before the DEER pulse is applied.
$\mu_0$ is the vacuum permeability.
$\sigma_j$ represents a relative change of the spin magnetic moment ($\langle S_z \rangle$) due to the spin flip between $|1\rangle \leftrightarrow |2\rangle$ transition of the $j$-th B spin, i.e., $\sigma_j = \pm |\sigma|$ where $|\sigma| = |\langle S_z \rangle_1 - \langle S_z \rangle_2|/2$. 
For example, $|\sigma| = 1/2$ for a simple $S=1/2$ spin.
$\vec{r}_j$ is the radius vector of the dipole interaction between the $j$-th B spin and the A spin.
$b_j^{MW} = b_j \left[ \delta_j^2 + \Omega^2 (\cos^2(\Omega_{B,j} t_p/2) - \sin^2(\Omega_{B,j} t_p/2))\right]/\Omega_{B,j}^2$ is the dipole magnetic field after the application of the DEER pulse.
$\Omega_{B,j} = \sqrt{\delta_j^2 + \Omega^2}$ where $\delta_j = \omega_{B} -\omega_j$ and
$\omega_j$ is the Larmor frequency of the $j$-th B spin.
With $\delta \varphi$, as shown previously~\cite{stepanov2016determination}, the intensity of the normalized DEER signal ($I = \langle
\cos(\delta\varphi)\rangle$) can be calculated as,
\begin{equation}
\label{eq:b}
I(n, \Delta \omega, \omega_B) =
\exp \left(- \frac{4 \pi \mu_0 \mu_B^2 g_A g_B T |\sigma|}{9\sqrt{3}\hbar} P_B(\Delta \omega, \omega_B) n \right),
\end{equation}
where $T$($\sim\tau_{1}$) is the delay between probe $\pi/2$ and DEER pulse and $\omega_B$ is the frequency of the DEER pulse. $n$ is the spin concentration in units of the number of spins per cubic meter. $P_B(\Delta \omega, \omega_B)$ represents the effective population transfer of B spins with the DEER pulse~\cite{stepanov2016determination}, namely,
\begin{equation}
\begin{split}
\label{eq:v}
& P_B(\Delta \omega, \omega_B) = \int\displaylimits_{- \infty}^{+ \infty}
\frac{\Omega^2}{(\xi - \omega_B)^2 + \Omega^2} \\
& \times \sin^2 \left( \sqrt{(\xi - \omega_B)^2 + \Omega^2} \frac{t_p}{2} \right) L(\xi, \omega_0, \Delta \omega)\, \text{d}\xi,
\end{split}
\end{equation}
where $\omega_0$ is the Larmor frequency and $\Omega$ is the resonant Rabi frequency of B spins and $L(\xi, \omega_0, \Delta \omega)$ represents the ESR spectrum of B spins with the resonance frequency $\omega_0$ and linewidth of $\Delta \omega$.
In the present analysis, we considered the Lorentzian function of $L(\xi, \omega_0, \Delta \omega)$ for P1 ESR spectra and homogeneous $\Omega$, where $\Omega$ was determined from a measurement of the Rabi frequency of P1 centers.
Then, the spin concentration of the P1 centers was obtained by fitting Eq.~(\ref{eq:b}) to the DEER signal where $n$ and $\Delta \omega$ were fitting parameters.

\section{Methods}
The DEER experiment requires applications of two distinct microwave pulses where one is to manipulate the NV$^{-}$ probe spins, and the other is to excite target spins whose concentration is being determined.
This is accomplished with a microwave circuit with two independent microwave transmission lines.
As shown in Fig.\,\ref{fig:diagram}, the microwave circuit consists of two sets of microwave sources (SG38x, Stanford Research Systems, indicated as MW1 and 2 in Fig.\,\ref{fig:diagram}), switches, microwave amplifiers and microwave transmission lines placed on the diamond sample.
The transmission-line wires with a  $\sim 200\,\mu$m  diameter are positioned in parallel with a gap of $\sim 200\,\mu$m.
The excitation laser beam is focused in the middle of the gap between the transmission-line wires.

\begin{figure}[t]
\includegraphics[width = 55 mm]{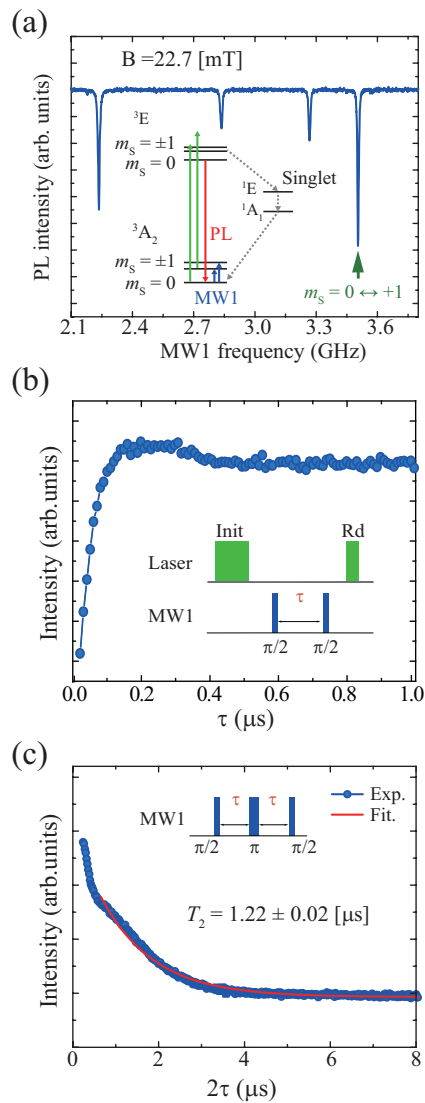}
\caption{\label{fig:cwODMRSE} ODMR experiment on a spot of the S5 sample (denoted as S5-1).
(a) cw ODMR spectrum of S5-1 taken at $B=22.7$ mT.
The field was aligned along the [111] axis within $\pm$1 degree.
The $\ket{m_S=+0} \leftrightarrow \ket{m_S=+1}$ transition used for the DEER experiment is indicated in this spectrum.
The inset shows the energy levels and transitions of the NV$^{-}$ centers in diamond.
(b) Ramsey signal of the NV$^{-}$ centers at S5-1.
Each point was obtained with an average of $\sim2,000$ measurements.
(c) Spin-echo decay signal of the NV$^{-}$ centers at S5-1.
The inset indicates the pulse sequence where the lengths of the $\pi/2$ and $\pi$ pulses are 90 and 180 ns, respectively.
}
\end{figure}
The ground state of the NV$^{-}$ center is a spin-triplet ($S=1$) manifold, see inset in Fig.\,\ref{fig:cwODMRSE} (a), where, in the absence of electric and magnetic fields or strain, the $\ket{m_S=0}$ sublevel lies 2.87$\ $GHz below the degenerate $\ket{m_S=\pm1}$ sublevels. %at room temperature.
Green laser light ($\lambda$ = 532 nm) with sufficient intensity pumps the majority of the NV$^{-}$ centers to $\ket{m_S=0}$, which is a ``bright state'' because the center cycles between the ground $^3$A$_2$ and the excited $^3$E electronic states under green-light illumination, emitting many PL photons in the process.
The ground-manifold states can be manipulated by the application of microwave fields.
When the population is transferred to the $\ket{m_S=\pm1}$ sublevels, the PL intensity decreases on account of these states being ``dark:'' once excited to the $^3$E state, the $\ket{m_S=\pm1}$ centers have significant probability of undergoing an intersystem crossing to the singlet states, where they do not participate in the excitation/PL cycle, effectively ``hiding'' in the metastable $^1$E state before eventually relaxing back to the $^3$A$_2$ state.
These properties enable optical detection of electron spin resonance (i.e., ODMR) in the ground manifold used here.

Figure~\ref{fig:cwODMRSE} shows measurement results of an ODMR experiment performed at a spot on the surface of the S5 sample (indicated as S5-1).
In this experiment, we align an external static magnetic field $\vec{B}$ to be along the [111] direction of the diamond lifting the degeneracy between the $\ket{m_S=\pm1}$ sublevels due to the Zeeman effect.
In this case, four distinct peaks in the ODMR spectrum are observed as shown in Fig.\,\ref{fig:cwODMRSE} (a).
This spectrum was taken by continuously illuminating the sample and scanning the microwaves [continuous-wave (cw) ODMR].
The two inner peaks correspond to the NV$^{-}$ centers oriented along the three diagonals of the cubic cell of the crystal that are at an angle to the magnetic field, while the outer peaks arise from the $\ket{m_S=0} \leftrightarrow \ket{m_S=-1}$ and $\ket{m_S=0} \leftrightarrow \ket{m_S=+1}$ transitions for the NV$^{-}$ centers oriented along the fourth diagonal which is collinear with the applied magnetic field.
From those four peak positions, the magnetic field strength was determined to be 22.7 mT, and its misalignment from the [111] axis was estimated to be within $\pm$ 1 degree.
We use the $\ket{m_S=0} \leftrightarrow \ket{m_S=+1}$ transition of the NV$^{-}$ spins in the pulsed ODMR experiment described below.

The basic pulsed ODMR sequences are those of a free-induction decay (FID) and spin-echo (SE) experiment~\cite{hahn1950spin} as depicted in the inset of Fig.\,\ref{fig:cwODMRSE}\,(b) and (c).
After initialization with a light pulse, a $\pi$/2 microwave pulse (where the terminology stems from an analogy between a general spin-1/2 two-level system) produces a coherent superposition of the $\ket{m_S=0}$ and $\ket{m_S=+1}$ states.
The excitation is followed by free evolution for a length of $\tau$ during which different NVs dephase from each other with a rate of $1/T_2^*$ due to the differences in their local magnetic environment.
For the FID experiment, to probe the state of the spins after the evolution, a second $\pi$/2 pulse is applied followed by a second light pulse and detection of the PL intensity [the inset of Fig.\,\ref{fig:cwODMRSE}\,(b)].
For the SE experiment, after a time interval of $\tau$, a $\pi$ pulse is applied, in turn followed by free evolution for another time interval $\tau$.
The spin state is then measured via the PL intensity after the application of a second $\pi$/2 pulse.
The echo scheme removes the effect of (quasi) static differences in the magnetic environments and allows to probe ``homogeneous'' $T_{2}$ relaxation in the presence of much faster ``inhomogeneous'' relaxation characterized by $T_{2}^*$.
The FID and SE signals for S5-1 are shown in the main part of Fig.\,\ref{fig:cwODMRSE}\,(b) and (c).
The decoherence time ($T_2$) is estimated to be $T_{2} = 1.22$ $\mu$s by fitting the curve with a single exponential function [see Fig.\,\ref{fig:cwODMRSE}\,(c)].

\begin{figure*}[t]
\begin{center}
\includegraphics[width=13 cm, clip]{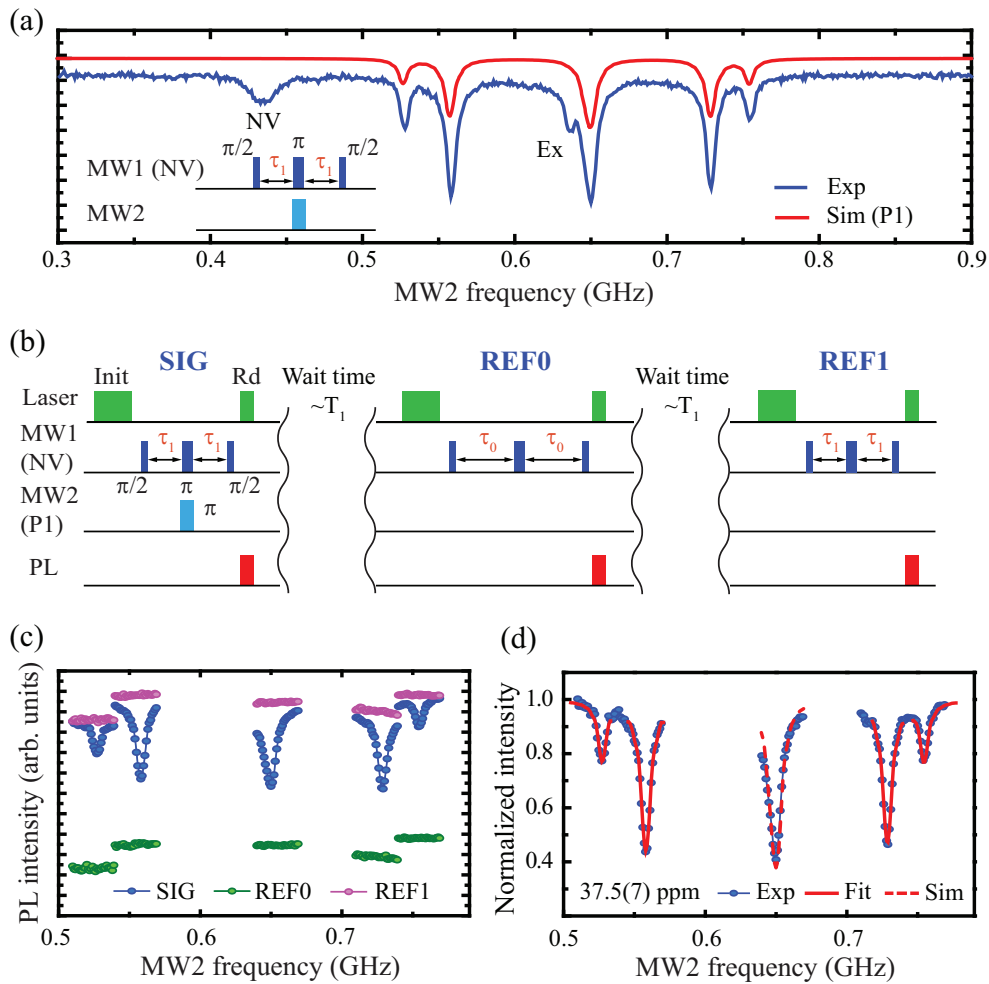}
\caption{\label{fig:DEERPulse} 
Overview of the DEER experiment. 
(a) DEER measurement of the S5-1 spot. 
The pulse sequence for MW1 and MW2 is shown in the inset.
The measurement was performed at $B=22.7$ mT.
The red solid line shows the simulated ESR spectrum of P1 spins.
The two other extra paramagnetic spins are indicated by ``NV" and ``Ex".
(b) Pulse sequence for the DEER measurement.
For the DEER measurement, in addition to the SE sequence for the NV$^{-}$ centers, a DEER pulse ($\pi$ pulse for P1 centers) is applied.
The three-part sequence separated by a time duration of $\sim10$ ms ($\sim T_1$ of P1 centers) acquires two references [see Fig.\,\ref{fig:cwODMRSE}\,(c)] and the DEER signal.
Data from each of the three parts are averaged independently.
(c) PL intensities of S5-1 measured using the SIG, REF0 and REF1 pulse sequences.
In the measurement, $\tau_0$ and $\tau_1$ were 4\,$\mu$s and 700\,ns, respectively.
(d) The normalized PL intensity of the S5-1 spot. The red solid line represents the fit result using Eq.\,(\ref{eq:b}).
The simulated center peak based on the result of the 4-peak fit is indicated by the red dashed line.
}
\end{center}
\end{figure*}
Next, we perform ESR spectroscopy of P1 centers using a DEER sequence.
As shown in the inset of Fig.\,\ref{fig:DEERPulse}\,(a), the DEER measurement involves a SE measurement of NV$^{-}$ centers with a fixed evolution time of $\tau_{1}$ (MW1) but with an additional DEER pulse (MW2) applied to drive the spins of the magnetic impurity of interest, i.e., P1 spins in the present case.
Flipping the impurity spins perturbs the NV$^{-}$ centers via dipole-dipole interaction, resulting in additional relaxation for NV$^{-}$ centers.
In the experiment, changes in the spin relaxation rate are detected by measuring changes of the PL intensity representing the SE signal intensity of the NV$^{-}$ centers.
In particular, we record the data of the PL change as a function of the MW2 frequency.
Since the flipping of the target spins happens only when the MW2 frequency matches their ESR condition,
the DEER spectrum represents the ESR spectrum of the target spins.
Similar to conventional ESR spectroscopy, the intensity and width of the DEER spectrum are related to the concentration of the target spins.
Therefore, we can determine the P1 concentration by analyzing the DEER spectrum (see Sect.~\ref{sect:theory}).
Figure~\ref{fig:DEERPulse}(a) shows the result of DEER experiment.
The obtained DEER spectrum contains seven pronounced signals. 
As shown in Fig.\,\ref{fig:DEERPulse}\,(a), we found that five peaks among the signals agree with simulated ESR spectrum using the spin Hamiltonian of P1 centers \cite{Loubser_1978, Cook_1966, Takahashi08}, and the other two extra peaks are possibly other types of paramagnetic spins.
Thus, the result shows the NV centers couple to P1 centers and the other paramagnetic spins.
The spin concentration analysis requires a quantitative analysis of the ESR spectrum.
However, because of the frequency dependence of the microwave circuits, the pulse length of the $\pi$ pulse for each peak may be different.
The measurements of the five P1 signals (SIG) were therefore performed separately using a correct length of the $\pi$ pulse for each peak.
It is worth to notice that, in conventional nuclear magnetic resonance (NMR) or ESR spectroscopy, generally, the SE signal is recorded as a time-domain signal where the SE intensity can be determined by taking the maximum value or the integral of the SE signal.
Thus, the ``zero" intensity can be defined when no signal is observed.
However, in the ODMR measurement, a change of the SE intensity is measured as a relative change of the PL signal where the PL signal is nonzero even though the SE intensity is zero.
This makes it challenging to evaluate the absolute change from the ``zero" intensity in an ODMR measurement.
In this experiment we determine the zero intensity of the SE signal (REF0) by measuring the PL intensity with an evolution time ($\tau_0$) much longer than $T_2$.
In addition we also record the SE intensity with the same evolution time as the DEER measurement ($\tau_1$), but without the DEER pulse (REF1).
Figure~\ref{fig:DEERPulse}\,(b) shows a pulse sequence, consisting of the SIG, REF0 and REF1 sections to determine these signals in the same sequence, and each section is separated by a wait time of $\sim T_1$ to avoid the saturation of the P1 ESR transition.
Figure~\ref{fig:DEERPulse}\,(c) shows the results of a DEER experiment using the pulse sequence shown in Fig.\,\ref{fig:DEERPulse}\,(b), consisting of the signal (SIG) and references (REF0 and REF1) for five P1 ESR peaks.
Using those SIG, REF0 and REF1, the normalized DEER signal ($I_{DEER}$) is then obtained with
\begin{equation}
\label{Norm}
I_{DEER} = \frac{\textrm{SIG}-\textrm{REF0}}{\textrm{REF1}-\textrm{REF0}}.
\end{equation}
%Figure~\ref{fig:DEERPulse}(c) shows the SIG, REF0 and REF1 data of the S5-1 sample.
The DEER signal normalized using Eq.~(\ref{Norm}) is shown in Fig.\,\ref{fig:DEERPulse}\,(d).
We use this signal to determine the P1 spin concentration.

\section{Results and analysis} \label{analysis}
Here we discuss applications of DEER to study local paramagnetic spin contents in diamond. 
Local spin concentrations of paramagnetic centers were measured at multiple locations of the diamond samples.

\subsection{Determination of [P1]}
As shown in Fig.\,\ref{fig:DEERPulse}\,(d), the DEER data obtained from S5-1 show five pronounced peaks, i.e., signals at 527.1, 557.8, 649.7, 729.0, and 754.4 MHz, corresponding to ESR signals of P1 centers in diamond ($S = 1/2$ and $I = 1$).
The concentration of P1 spins was extracted by fitting the DEER data with Eq.~(\ref{eq:b}).
For S5-1, the fits were performed with four DEER peaks at 527.1, 557.8, 729.0 and 754.4 MHz and excluding the central peak at 649.7 MHz.
This is because 649.7 MHz in this experiment corresponds to the Larmor frequency for $g \sim 2$ spins, where there are often multiple signals, which may contribute as a systematic error in the analysis.
From the concentration analysis, we found the P1 spin concentrations to be 37.5(7) ppm for location 1 in the sample S5 (S5-1).
To estimate an average flip rate of P1 centers, we also analyzed the SE result of S5-1 in Fig.\,\ref{fig:cwODMRSE}\,(c) to extract the most probable spin-flip rate ($W_{max}$).
From the comparison between the SE data and the model [Eq.~(\ref{eq:a})], we determined $W_{max}=11\times 10^3$ $s^{-1}$  for [P1] = 37 ppm experimentally.
%Because a fast spin flip rate reduces the DEER signal intensity, the true [P1] should not be less than the experimentally determined [P1], and $W_{max}$ should also not be more than the experimentally determined value ($W_{max, exp}$).
%Therefore, $W_{max, exp}$ is a good indicator to check whether a significant contribution of the spin-flip events in the DEER experiment.
In the present case, with 2$\tau=1.4$ $\mu$s, we obtained the probability of spin flip events (2$\tau \langle W \rangle \sim 14.2 \tau W_{max}$) to be 0.11.
With the obtained small flip probability on the time scale of the DEER experiment, we consider the P1 spins to be static to model the DEER signal. This was the case for all DEER measurements.

\subsection{Determination of [Ex]}
\begin{figure}[t]
\begin{center}
\includegraphics[width = 7 cm]{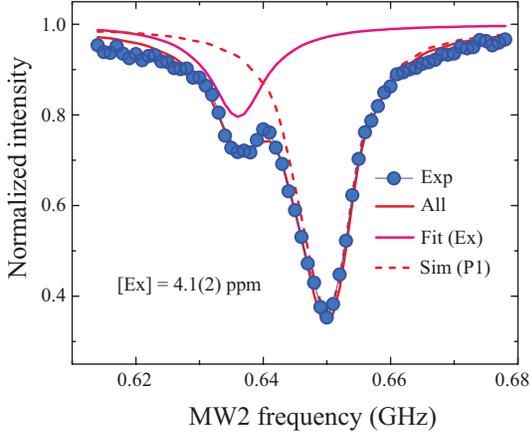}
\caption{\label{fig:S5DEER_Ex} Determination of [Ex]. The blue dots are experimental data and the pink solid line is the result of the analysis for Ex, red dashed line is the simulated spectrum of center peak of P1 centers, red solid line is the summation of fit of Ex and simulated spectrum of P1 centers. In the experiment, the length of the DEER pulse was set to be $\Omega t_p = \pi$.  
}
\end{center}
\end{figure}
In addition to ESR signals from P1 centers, two extra peaks were observed as shown in Fig.~\ref{fig:DEERPulse}(a).
One was observed at an ESR frequency for a $g \sim 2 $ spin ($\approx 0.64$ GHz) while the other was observed at $\approx 0.43$ GHz.
We labeled the $g \sim 2$ peak as ``Ex'' (see Fig.~\ref{fig:DEERPulse}(a)).
As shown in Fig.~\ref{fig:S5DEER_Ex}, we obtained [Ex] to be 4.1(2) ppm from the best fit using Eq.~(\ref{eq:b}). 
In the analysis we assumed that the signal is originated from an $S = 1/2$ and $g = 2$ spin.
Since Ex peak partially overlaped with the center peak of P1 centers, the Ex peak was extracted by subtracting the P1 spectrum as seen in Fig.~\ref{fig:S5DEER_Ex}.

\subsection{Determination of [NV$^{-}$]}
\begin{figure*}[t]
\begin{center}
\includegraphics[width = 16 cm, clip]{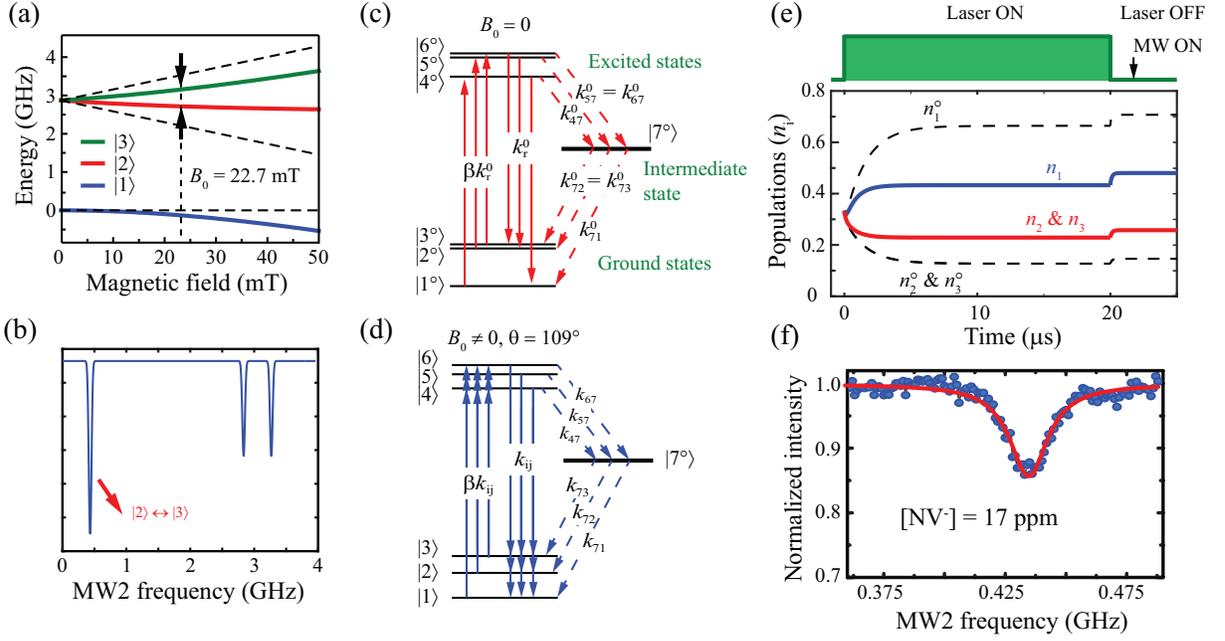}
\caption{\label{fig:S5DEER_NV} 
(a) Energy levels of the NV$^{-}$ centers as a function of $B_0$.
The blue, red and green solid lines represent the energy levels when the magnetic field orientation is 109.5 degrees away from the NV$^{-}$ axis ($\theta = 109.5$), corresponding to the $[\bar{1}11$], $[1\bar{1}1$] and $[11\bar{1}$] orientations.
The black dashed lines represent the energy levels with the magnetic field along the NV$^{-}$ axis. 
(b) The simulated DEER signals. 
The simulation assumed that $\pi$-pulses were applied for the DEER measurement, namely $\Omega t_p = \pi$.
Energy diagrams and transition rates at $B_0 = 0$ (c) and $B_0 \ne 0$ (d). 
In the present study, we assume that the intermediate state is unchanged under an application of magnetic field.
(e) Calculation of the populations of the states and a sequence of the excitation laser. In the present case, the length of the excitation laser is 20 $\mu$s and $B_0 = 22.7$ mT. 
The polar angle ($\theta$) is 109.47 degrees. The result is independent of the azimuthal angle. $\beta = 0.03$, $k^0_r = 63.7$ $\mu$s$^{-1}$, $k^0_{47} = 9.6$ $\mu$s$^{-1}$, $k^0_{57} = 53.0$ $\mu$s$^{-1}$, $k^0_{71} = 0.9$ $\mu$s$^{-1}$ and $k^0_{72} = 0.5$ $\mu$s$^{-1}$ \cite{robledo2011spin}. 
The dashed lines are the populations at $B_0 = 0$ ($n^0_1$, $n^0_2$ and $n^0_3$). 
(f) Analysis of [NV$^{-}$]. The blue dots are experimental data and the red solid line is the result of the analysis. In the experiment, the length of the DEER pulse was set to be $\Omega t_p = \pi$.
}
\end{center}
\end{figure*}
Furthermore, we identified the peak at 0.43 GHz to be a NV transition based on the spectral analysis and demonstrated the determination of [NV$^{-}$]. 
Here we describe the method to determine [NV$^{-}$] using the NV$^{-}$-detected ESR spectrum depicted in Fig.\,\ref{fig:DEERPulse}\,(a). 
%The observed ESR signal is originated from the off-axis NV centers, which is suitable for the determination of [NV$^{-}$]. 
First, we explain the origin of the ESR signal using the NV spin Hamiltonian at a given magnetic field, and then, discuss the effect of the spin polarization due to the optical excitation of the NV center, with which we find that the polarization of the observed ESR transition is negligible. Finally, we show the result of the [NV$^{-}$] analysis.
The spin Hamiltonian of the NV$^{-}$ center is given by,
\begin{equation}
   H = D S_z^2 + \frac{g \mu_B}{h} \mathbf{B_0} \cdot \mathbf{S}
   \label{eq:HNV}
\end{equation}
where $\mathbf{S} = (S_x, S_y, S_z)$ are spin operators for NV centers, $g$ is the $g$-factor ($g$ = 2.0028), $h$ is the Planck constant, $\mu_B$ is the Bohr magneton, $\mathbf{B_0}$ is the external magnetic field vector, and $D = 2.87$ GHz is the zero-field splitting constant of the NV ground state. 
Figure \,\ref{fig:S5DEER_NV}\,(a) shows the calculated energy levels of the NV$^{-}$ center as a function of the magnetic field strength. 
The calculation was done using Eq.~(\ref{eq:HNV}). 
In the calculation, the angle between the magnetic field and the NV axis was set to be $109.5$ degrees, corresponding to the angle between $[111]$ and $[\bar{1}11]$ (also $[1\bar{1}1]$ and $[11\bar{1}]$) orientations. 
As shown in Fig.\,\ref{fig:S5DEER_NV}\,(a), the NV$^{-}$ center has three energy levels, labeled $|1\rangle$, $|2\rangle$ and $|3\rangle$. 
The energy gap between the $|2\rangle$ and $|3\rangle$ states at a magnetic field of 22.7 mT is also indicated ($\Delta E = 0.432$ GHz). 
It is worth noting that the transition probability between $|2\rangle$ and $|3\rangle$ under microwave excitation ($k_{12}$) becomes finite when $B_0 \ne 0$ because both $|2\rangle$ and $|3\rangle$ states consist of superpositions of $|m_S = 1\rangle$, $|m_S = -1\rangle$ and $|m_S = 0\rangle$. 
Using this condition, we calculated the DEER spectrum of the  NV$^{-}$ centers. 
The simulated DEER spectrum of the NV$^{-}$ center is shown in Fig.\,\ref{fig:S5DEER_NV}\,(b).
The calculation was done using Eq.\,(\ref{eq:b}). 
As discussed, the intensity of the DEER signal depends on $|\sigma|$.
In the present case, $|\sigma| = |\langle 2 | S_z | 2\rangle - \langle 3 | S_z |3 \rangle | = 0.94$ is approximately twice that for a regular ESR transition ($|\sigma|=1/2$). 
Therefore, the $|2\rangle \leftrightarrow |3\rangle$ transition is more pronounced than the other peaks.

Next we calculate the population of all states. 
A linear combination of the zero-field eigenstates is expressed by,
\begin{equation}
| i \rangle = \sum_{j=1}^7 \alpha_{ij} |j^0 \rangle,
\end{equation}
where $|a_{ij}|^2$ represents the population of $| j \rangle$-state in $| i \rangle$-state. 
$a_{ij}$ is numerically calculated with the spin Hamiltonian. The calculation of the ground state Hamiltonian in Eq.~(\ref{eq:HNV}) determines ${a_{ij}}$ for $i,j = 1,2$ and 3. 
Similarly, the calculation of the excited states Hamiltonian (Eq.~(\ref{eq:HNV}) with $D \xrightarrow{} D_e = 1.42$ GHz) determines ${a_{ij}}$ for $i,j = 4,5$ and 6. 
$a_{77} = 1$ in the present case. 
The rest of ${a_{ij}}$ are zero. 
Transition probability for the $|i \rangle \leftrightarrow |j \rangle$ transition is calculated by the square of its electric dipole transition moment, {\it i.e.} $| \langle i| \mu | j \rangle |^2$, where $\mu$ is the electric dipole moment operator. 
Since all transitions are assisted by phonon couplings~\cite{goldman2015phonon}, the transitions are considered to be incoherent transitions~\cite{tetienne2012magnetic}. 
Therefore, we model the transition rates of those transitions ($k_{ij}$) to be a sum of the zero-field transition rates $(k_{pq}^0 = | \langle q^0 | \mu | p^0 \rangle |^2$ where ${|p^0 \rangle }$ are the zero-field states) weighted by the populations $|a_{ip}|^2$ and $|a_{jq}|^2$, namely,
\begin{equation}
\begin{split}
k_{ij} = & |\langle j | \mu | i \rangle |^2 = \sum_{p=1}^7 \sum_{q=1}^7 |a_{ip} a^*_{jq}|^2 |\langle q^0 |\mu| p^0 \rangle|^2 \\
& = \sum_{p=1}^7 \sum_{q=1}^7 |a_{ip}|^2 |a_{jq}|^2 k_{pq}^0.
\end{split}
\end{equation}
Finally, the population of the NV state ($n_i$) is calculated using a system of the rate equations,
\begin{equation}
\frac{dn_i}{dt} = \sum_{j=1}^7 (k_{ji} n_j - k_{ij} n_i), 
\label{eq:rate}
\end{equation}
where $\Sigma_i n_i = 1$. 

Figure \,\ref{fig:S5DEER_NV}\,(e) shows transient populations of NV states during the DEER experiment. 
The populations were numerically calculated using Eqs.~(\ref{eq:HNV})-(\ref{eq:rate}). 
As shown in Fig.\,\ref{fig:S5DEER_NV}\,(e), in the DEER experiment the laser excitation with a length of 20 $\mu$s was used and the wait time with a length of $\sim 2.5$ $\mu$s was applied after the laser excitation to initialize the NV spin state before microwave excitations for the experiment. 
As shown in Fig.\,\ref{fig:S5DEER_NV}\,(e), the populations of the ground states reached an equilibrium under the 20 $\mu$s laser excitation, and then the populations were built up in the ground states. 
As can be seen in the figure, we found that the population of $n_2$ and $n_3$ were 26 $\%$ and $n_1$ was 48 $\%$ after the wait time. This corresponds to the spin polarization $[(n_1 - n_2)/(n_1 + n_2)]$ of 30 $\%$ for the $|1\rangle \leftrightarrow |2\rangle$ transition. 
Compared with the case at $B = 0$ ($n_{02}$ and $n_{03} = 15$ $\%$ and $n_{01} = 70$ $\%$, the spin polarization of 65 $\%$), the spin polarization at $B_0 \ne 0$ is much smaller than that at $B_0 = 0$. Furthermore, since $n_2$ and $n_3$ are same, the result verified the unpolarized condition of the $|2\rangle$ and $|3\rangle$ transition, which is used in the concentration analysis with the present DEER model, and estimated that 52 $\%$ of the off-axis NV centers are involved in the DEER signal. 

Finally, we discuss the analysis of [NV$^{-}$]. Figure~\,\ref{fig:S5DEER_NV}\,(f) shows the normalized DEER signal. 
As described above, the obtained DEER signal is originated from 52 $\%$ of the off-axis NV centers consisting of the $[\bar{1}11]$,$[1\bar{1}1]$ and $[11\bar{1}]$ orientations.
By taking into account those and using Eq.~(\ref{eq:b}), we obtained [NV$^{-}$] to be 17 ppm.
We decided not to provide the uncertainty of [NV$^{-}$] because of the model-dependence in the determination.
% estimated the concentration of NV spins to be 17.5(6)\,ppm from the DEER analysis of this signal.
%The concentration was determined assuming no net spin polarization of the off-axis NV centers. 
%Furthermore, using the same analysis described in this section, we extracted the spin concentration [Ex] to be 4.1(2)\,ppm.
%In the analysis we assume that the origin of the ``Ex'' signal is a $g=2$ and $S=1/2$ spin.

\subsection{Study of local spin contents at various locations}
\begin{figure*}[t]
\begin{center}
\includegraphics[width = 15 cm, clip]{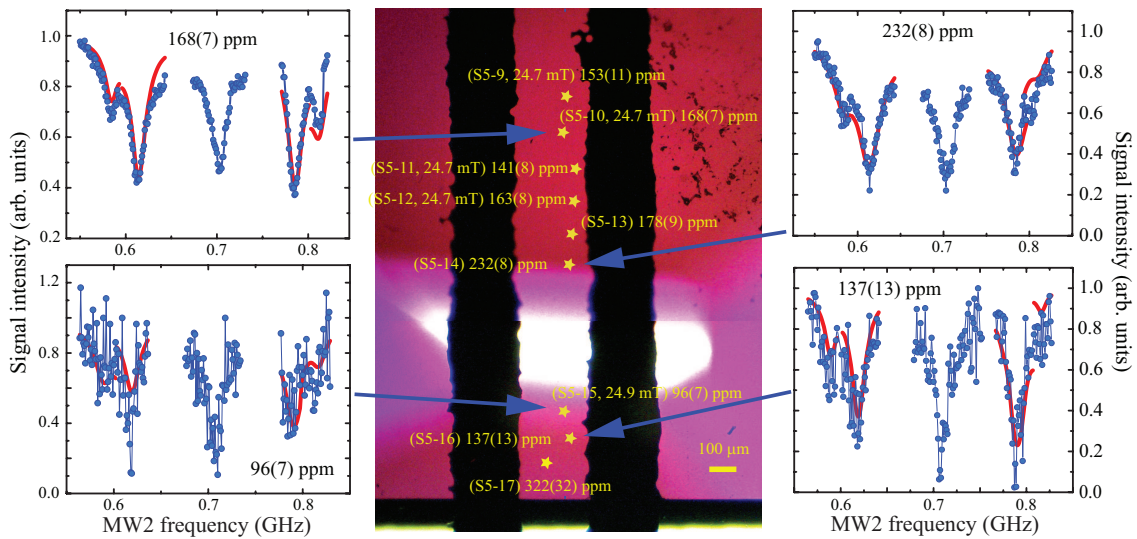}
\caption{\label{fig:S5DEER} The local P1 spin concentrations of the S5 sample.
The determined spin concentrations and their locations are overlayed on an optical microscope image of the S5 sample.
The concentrations are obtained from the DEER analysis.
The measurements from S5-5 to S5-10 were performed at $B=24.7$ mT.
The magnetic fields for the measurements of S5-11, 12 and 13 were 24.9 mT, 24.9 mT, and 31.7 mT, respectively.
The DEER data and the analysis results at selected locations are shown in the right and left sides of the figure. The white region is transparent and free from [NV$^-$]. The purple region including the spot S5-1 gives bright fluorescence.  
}
\end{center}
\end{figure*}
We studied various locations on the S5 sample.
The measurement locations were chosen for the study to cover the surface uniformly.
Through the measurements, we found that the spin concentrations are similar in some locations (for example, around the S5-1 spot) and inhomogeneous in other locations.
Figure~\ref{fig:S5DEER} shows the results from the location with a highly inhomogeneous spin concentration.
The results were obtained at locations over several hundred micrometers along the microwave wires.
As shown in Fig.\,\ref{fig:S5DEER}, we observed that the spin concentrations of P1 centers on the S5 sample surface range from 96 ppm to 322 ppm.

\begin{figure*}[t]
\begin{center}
\includegraphics[width = 18 cm]{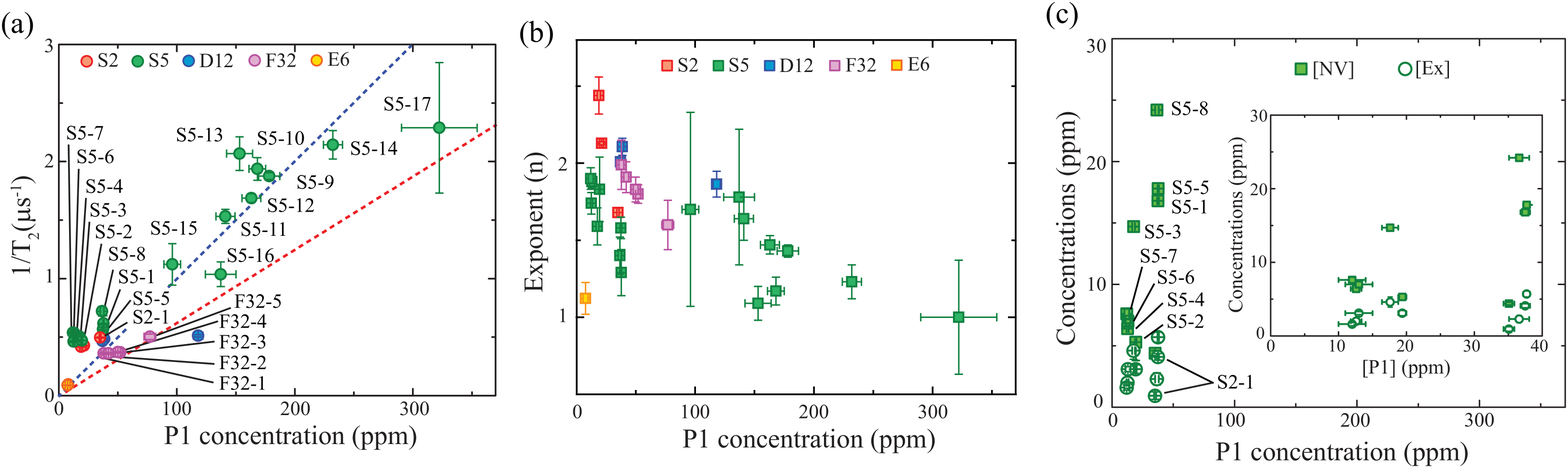}
\caption{\label{fig:T2vsConc}
Summary of the extracted relaxation rates and P1, NV$^-$ and Ex spin concentrations.
(a) $1/T_2$ as a function of the P1 concentration for all measured samples. The red dotted line represents $1/T_2$ ($\mu$s$^{-1}$) $= 1/160$ ($\mu$s$^{-1}$ ppm$^{-1}$)$\times$[P1] (ppm) found in Ref.~\cite{bauch2020decoherence}.
The blue dotted line represents $1/T_2$ ($\mu$s$^{-1}$) $= 1/100$ ($\mu$s$^{-1}$ ppm$^{-1}$)$\times$[P1] (ppm).
The SE decay envelope was fitted to the stretched exponential function ($\propto \exp[-(t/T_2)^n]$) to obtain $T_2$. 
(b) The exponents (n) as a function of [P1].
(c) NV and Ex spin concentrations as a function of P1 concentration.
The inset shows the NV and Ex spin concentrations in a magnified scale.
}
\end{center}
\end{figure*}
Finally, we summarize all analysis results from the S2, S5, D12, F32 and E6 samples.
Figure~\ref{fig:T2vsConc}\,(a) shows $1/T_2$ values of the NV$^{-}$ centers as a function of [P1].
The $T_2$ values were determined by fitting the SE decay envelope with the stretched exponential function $A \exp[-(t/T_2)^n$].
As can be seen in Fig.~\ref{fig:T2vsConc}\,(a), the P1 concentrations widely vary among the samples.
The average P1 concentrations are 25 ppm, 120 ppm, 64 ppm, and 52 ppm for the S2, S5, D12, and F32 samples, respectively.
In addition, a measurable inhomogeneity in the P1 concentration was observed for all samples.
For example, the P1 concentration in the S5 sample ranges between 13 ppm and 322 ppm, which is a variation by more than an order of magnitude.
Furthermore, a linear relation between $1/T_2$ values and [P1] was reported by the previous studies of $T_2$ in diamond~\cite{vanWyk97, wang2013spin, Takahashi08, stepanov2016determination,bauch2020decoherence} which investigated a wide range of P1 concentrations (from tens of ppb to hundreds of ppm).
As shown in Fig.\,\ref{fig:T2vsConc}\,(a), it seems there are also some correlations between $1/T_2$ values and P1 concentrations in the present study.
However, a linear dependence is much less clear than in previous studies.
A linear function of $1/T_2$ ($\mu$s$^{-1}) = 1/160$ ($\mu$s$^{-1}$ ppm$^{-1}$)$\times$[P1](ppm) found in Ref.~\cite{bauch2020decoherence} is shown as a red dotted line in Fig.~\ref{fig:T2vsConc}\,(a).
Most of the obtained $1/T_2$ values are located above the dotted line.
A possible reason for the observation is a significant contribution of the $T_2$ relaxation process from other paramagnetic impurities, including neighboring NV$^{-}$ centers. 
Figure~\ref{fig:T2vsConc}\,(b) shows the obtained exponents as a function of [P1]. 
As shown in the figure, the exponents, ranging between $\approx 2.5$ and $1$, are closer to $1$ at spots with high [P1].
The concentrations of NV centers and the defects responsible for the extra peaks are shown in Fig.\,\ref{fig:T2vsConc}\,(c).
As can be seen in the figure, the NV concentrations are larger than those of the $g=2$ spins and the correlation between the NV and P1 concentrations is unclear. For the S5 sample with strong nonuniformity of the irradiation dose, a direct correlation between these concentrations is not expected.
%\textbf{Dima says: regarding the last point, we do not really expect a clear correlation between [NV] and [P1] because S5 is an ``irradiation accident'' that has made it so non-uniform. I wonder if we should remove the part about: the correlation between the NV and P1 concentrations is unclear ? } 

%As can be seen from Fig.~\ref{fig:T2vsConc}, we found that no correlation between the sample location and $T_2$ from other impurities.
%The result suggests that the different conversion factor from P1 to NV during sample preparation from sample to sample and location to location.

\section{Summary and outlook}
In conclusion, we presented a local measurement technique used to determine the concentration of defects in diamond crystals utilizing double electron-electron resonance. 
We performed measurements on several different spots on different samples and determined locally the concentration of P1 centers in the crystals.
As we demonstrated to obtain [NV$^{-}$] and [Ex], this technique is applicable to various impurities spins in diamond as well as external spins located near the diamond surface.
%This technique should be applicable to different impurities other than P1 like NV$^{0}$, paramagnetic surface impurities as well as NV$^{-}$ centers. 
In the follow-up research, we plan to make systematic measurements at low temperature to study the dependence of $T_{1}$ on local impurity concentration in order to understand the mechanisms of the relaxation and explain the observations in Refs.~\cite{jarmola2012temperature, mrozek2015longitudinal}. 
Also, we envision applications of this technique in material characterization, controlled growth, and sensor optimization.

\begin{comment}
\textbf{Colleagues: this part is super important! Please think what we can add to this list...}
* Presented measurements are mostly on P1. They nicely complement bulk-ESR measurements
* Extension to other impurities
* In the follow-up research, we plan to make systematic measurements at low temperature to study the dependence of T1 on local impurity concentration in order to understand the mechanisms of the relaxation at explain the observations in Refs.\ \cite{jarmola2012temperature,mrozek2015longitudinal}.
* In the future, we envision application to material characterization, controlled growth, and sensor optimization
\end{comment}

\begin{acknowledgments}
The authors acknowledge helpful discussions with Yehuda Band, Connor Hart, Junichi Isoya, Kai-Mei Fu, Ron Walsworth, Norman Yao and Jean-Philippe Tetienne. DB acknowledges support from the AFOSR/DARPA QuASAR program at the early stages of this work, as well as support from the DFG through the DIP program (FO 703/2-1), the German Federal Ministry of Education and Research (BMBF) within the Quantumtechnologien program (FKZ 13N14439), and EU FET-OPEN Flagship Project ASTERIQS (action 820394).
ST acknowledges support from the National Science Foundation (DMR-1508661, CHE-1611134, and CHE-2004252), the USC Anton B. Burg Foundation, and the Searle Scholars Program. 
YS was supported by Grant-in-Aid for Scientific Research on Innovative Areas (No.19H05824). 
A.J. acknowledges support from the U.S. Army Research Laboratory under Cooperative Agreement No. W911NF-16-2-0008.

\end{acknowledgments}

\bibliography{Li}
\end{document}